\documentclass[conference]{IEEEtran}
\IEEEoverridecommandlockouts
\usepackage{cite}
\usepackage{amsmath,amssymb,amsfonts}
\usepackage{algorithm}
\usepackage{graphicx}
\usepackage{textcomp}
\usepackage{xcolor}
\usepackage{amsmath}
\usepackage{hyperref}
\usepackage{tikz}
\usetikzlibrary{matrix,shapes,arrows,positioning}
\usepackage{pifont}
\usepackage{xspace}
\usepackage{amsmath}
\usepackage{caption}
\usepackage{subcaption}
\usepackage{placeins}
\usepackage{balance}
\usepackage{cite}
\usepackage{bm}
\usepackage{amsmath}
\usepackage{array}
\usepackage{setspace} 
\usepackage[nameinlink]{cleveref}
\usepackage{accents}
\usepackage{mathtools}

\usepackage{amsthm} 
\usepackage{multirow}
\usepackage{tabularx}
\usepackage{color}
\usepackage{epstopdf}
\usepackage{endnotes}
\usepackage{framed}
\usepackage{subcaption}
\usepackage{cool} 
\usepackage{enumerate} 
\Crefname{figure}{Fig.}{Figs.}

\usepackage{algpseudocode}
\def\BibTeX{{\rm B\kern-.05em{\sc i\kern-.025em b}\kern-.08em
    T\kern-.1667em\lower.7ex\hbox{E}\kern-.125emX}}
\begin{document}

\title{Charging Ahead: A Hierarchical Adversarial Framework for Counteracting Advanced Cyber Threats in EV Charging Stations}
\author{\IEEEauthorblockN{Mohammed Al-Mehdhar,  Abdullatif Albaseer, Mohamed Abdallah, Ala Al-Fuqaha}
\IEEEauthorblockA{ Division of Information and Computing Technology, College of Science and Engineering, \\ Hamad Bin Khalifa University, Doha, Qatar \\
\{moal44567, aalbaseer, moabdallah, aalfuqaha\}@hbku.edu.qa}
}

\maketitle

\begin{abstract}
The increasing popularity of electric vehicles (EVs) necessitates robust defenses against sophisticated cyber threats. A significant challenge arises when EVs intentionally provide false information to gain higher charging priority, potentially causing grid instability. While various approaches have been proposed in existing literature to address this issue, they often overlook the possibility of attackers using advanced techniques like deep reinforcement learning (DRL) or other complex deep learning methods to achieve such attacks. In response to this, this paper introduces a hierarchical adversarial framework using DRL (HADRL), which effectively detects stealthy cyberattacks on EV charging stations, especially those leading to denial of charging. Our approach includes a dual approach, where the first scheme leverages DRL to develop advanced and stealthy attack methods that can bypass basic intrusion detection systems (IDS). Second, we implement a DRL-based scheme within the IDS at EV charging stations, aiming to detect and counter these sophisticated attacks. This scheme is trained with datasets created from the first scheme, resulting in a robust and efficient IDS. We evaluated the effectiveness of our framework against the recent literature approaches, and the results show that our IDS can accurately detect deceptive EVs with a low false alarm rate, even when confronted with attacks not represented in the training dataset.
\end{abstract}

\begin{IEEEkeywords}
Hierarchical Adversarial Reinforcement Learning,
Intrusion Detection Systems (IDS),
Electric Vehicle Charging Stations (EVCS),
State of Charge (SoC) Manipulation,
\end{IEEEkeywords}

\section{Introduction}
\label{sec:into}
The pursuit of smart city development, characterized by efficiency, reliability, sustainability, and interconnectivity, necessitates a shift towards electric vehicles (EVs) as the primary mode of future transportation. This transition aligns with the broader objectives of smart city initiatives, including reducing carbon emissions and lessening dependence on crude oil~\cite{energy_report, barman2023renewable}.  
The increasing use of EVs marks a notable transition in urban mobility patterns, further evidenced by a growing preference for environmentally sustainable transport modes~\cite{ismail2023impact}.

However, deploying EVs at scale introduces several challenges, including the necessity for charging coordination at EV charging stations (EVCS), which can serve only a limited number of cars at a given time~\cite{harnett2018doe}. Also, the concurrent and uncoordinated charging of EVs could threaten grid stability. 
Various strategies have been developed to coordinate the charging load with the available power supply, involving the communication of key information like the State of Charge (SoC) of the battery to a centralized charging coordinator (CC). However, the accuracy of the data provided by the EVs is a critical assumption in these mechanisms, and it is not always valid~\cite{elghanam2021review }. In addition, a critical concern beyond charging coordination is the security risk associated with integrating wireless technology (Wi-Fi, cellular, Bluetooth, etc.) in EVCS. These risks, including identity theft and advanced persistent threats (APT) such as ransomware and malware, position EVCS as potential entry points for cyber-attacks~\cite{harnett2018doe}.
Specifically, EVs can initiate a Distributed Denial of Service (DDoS) attack on charging stations by overwhelming the network with fraudulent requests \cite{antoun2020detailed}. These attacks can overwhelm the charging schedules and frustrate other vehicles from accessing the grid. Additionally, malicious EVs can alter the "charging profile" to amplify the demand on the grid during periods of high usage. The smart grid encounters challenges in meeting the demand of the connected load during such occurrences, potentially impeding the provision of power to legitimate consumers. Moreover, EVs may intentionally provide inaccurate data, such as lower SoC values, to ensure they can charge before their charging requests expire and earn higher priority in the charging schedule~\cite{alomrani2023detecting}.

Therefore, establishing a robust and secure EVCS infrastructure is crucial. For example, intrusion detection systems (IDSs), enhanced by advanced machine learning (ML) and deep learning (DL) algorithms, can provide user benefits like remote monitoring and the ability to schedule off-peak EV charging and detect misleading information~\cite{liu2023building, antoun2020detailed}. 
Building ML/DL-based IDS has gained significant attention in recent literature. This is primarily due to their enriched capability to efficiently detect malicious behavior, thereby remarkably improving overall system performance. For example, Basnet et al. \cite{basnet2020deep} introduced a DL-based IDS to identify potential DoS attacks. Specifically, the proposed approach integrates two distinct neural network (NN) architectures: the deep NN and the Long Short-Term Memory network (LSTM). This finding emphasizes the potential of LSTM networks to enhance the security framework of IoT-enabled EVCS against sophisticated cyber threats. In~\cite{shafee2020detection}, authors explored an ML  solution to identify malicious EVs that manipulate SoC data to mislead the CC. The authors applied a gated recurrent unit (GRU) architecture, using actual charging patterns of plug-in hybrid EVs and simulated false reporting attacks. 

However, those works \cite{basnet2020deep, shafee2020detection} used handcrafted attacks where they often fail to accurately replicate real-world attack scenarios, leading to a potential lack of generalizability in IDS when encountered with sophisticated attacks. In response, lately, the work in \cite{alomrani2023detecting} proposed using reinforcement learning (RL) to build a more challenging attack than the handcrafted ones.  Specifically, an RL model is introduced to generate stealthy attacks to train a better DL-based IDS. Nevertheless, some critical limitations still remain. First, one of the primary limitations is the inability to capture the complicated stealthy attack patterns effectively. As cyber threats evolve and become more sophisticated, a straightforward RL agent may need help adapting to and identifying these complex patterns, especially if the attacker uses a more advanced DL model architecture to generate stealthy attacks, such that the IDS's performance may suffer significantly, leading to a sharp degradation in its accuracy and overall effectiveness. This necessitates the need for a more reliable approach not only to generate and mimic more destructive stealthy attacks but also to develop a more robust DL-based IDS.

Motivated by these remarks, in this paper, we design and propose a framework that includes designing more destructive attacks and developing a more robust IDS at the EVCS. Our proposed framework consists of two schemes, each with a distinctive objective. The first scheme leverages the power of deep RL (DRL) to create sophisticated synthetic attacks. These attacks stand out for their attention to temporal relationships, adding a layer of complexity to their structure. In the second scheme, we implement a robust IDS based on DRL. This IDS leverages the complicated attack patterns generated by the adversarial DRL agent.
Finally, we showcase the resilience of our framework compared to the baselines (i.e., RL and manually crafted attacks).   
 

    
The remaining parts of this paper are structured as follows: In Section \ref{sec:sysmodel}, we present the system model and the problem formulation. Section \ref{sec:proposed_solution} clearly introduces the proposed solution, including the two proposed schemes for attack generation as well as for robust DRL-based IDS. We evaluate the proposed framework in Section \ref{sec:results} while we conclude the whole paper in Section \ref{sec:concl}.

\begin{figure}[t]\centering
\includegraphics[width=0.8\linewidth]{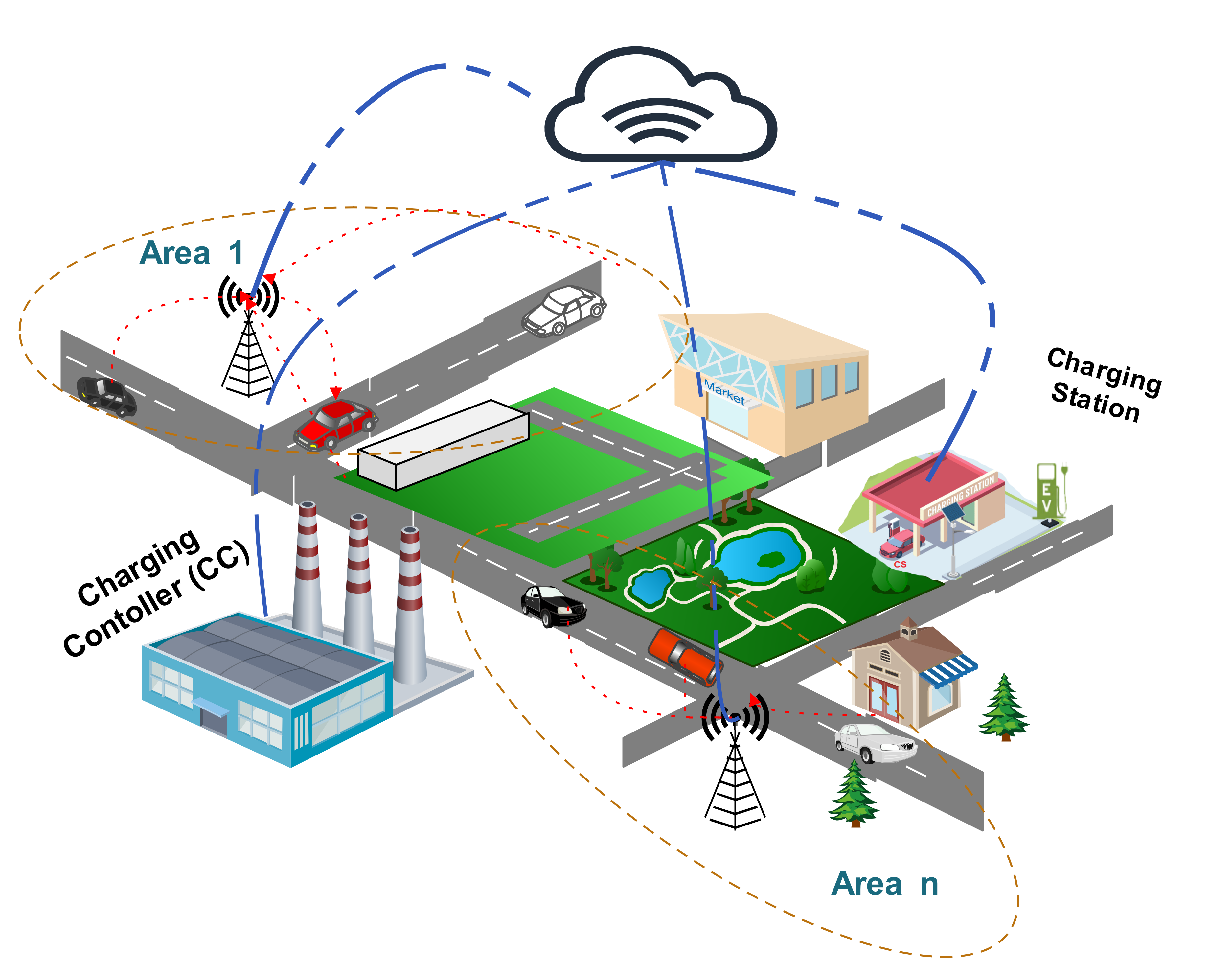}
\caption{The system model with ($K$) EVs, charging station, and charging controller.}

\label{fig:figure1}
\end{figure}

\section {System Model and problem formulation}
\label{sec:sysmodel}

\subsection{System Model}
As illustrated in \Cref{fig:figure1}, the system model encompasses the CC, denoted by \(i\)) and a set of EVs, denoted by \(\mathbb{K}: \{EV_1, EV_2, \ldots, EV_K\}\) where $K$ is the number of EVs and $k = 1, 2, \dots, K$ is the EV index. The model includes the following integral components, each with specific functionalities and interactions: \\
\textbf{ CC:} The CC serves as the central decision-making unit. It processes incoming charging requests from EVs and computes optimized charging schedules. These schedules are carefully designed to balance the charging needs of each EV with the station's current energy capacity and grid stability requirements. The CC also monitors real-time energy consumption and adjusts schedules dynamically to maintain efficiency and prevent grid overload. \\
\textbf{EVs:} Each EV communicates its charging requirements to the CC, including crucial battery SoC statistics and the expected charging duration. This data is essential for the CC to accurately allocate resources and prioritize charging schedules based on urgency and energy availability. EVs are also equipped with mechanisms to authenticate their communication with the CC, ensuring data integrity and security.\\
\textbf{Aggregator:} Acting as an intermediary, the Aggregator collates and forwards EV charging requests from a specific locale to the CC. Post-scheduling, it disseminates the CC’s directives back to the EVs. This entity plays a key role in reducing communication overhead and streamlining the data flow between numerous EVs and the CC, thus enhancing the overall efficiency of the charging process.\\  
\textbf{Charging Points (CPs):} CPs are the physical interface through which EVs receive electrical power for charging. They are equipped with advanced metering and control systems to regulate power supply according to the CC's schedules. CPs can also report real-time charging status and power quality metrics back to the CC, enabling proactive management of the charging infrastructure.

The core objective of this paper is to develop a sophisticated DL-based IDS adept at detecting and neutralizing covert cyberattacks that manipulate SoC data. Falsified SoC values primarily characterize these attacks in charging requests perpetrated by a malicious EV aiming to gain unwarranted charging priority. The attack model proposed in this study assumes advanced reconnaissance capabilities for the attacker, who can intercept and modify communications between EVs and EVCS. This model presupposes that the attacker has comprehensive access to incoming charging requests at the CC but is limited to manipulating the SoC data of their own EV. 
\vspace{-2pt}
\subsection{Probelm Formulation}
Given the set of EVs $\mathbb{K}$ defined above, a set of days \(\mathbb{D} = \{D_1, \ldots, D_d\}\), and a set of equal time slots \(T = \{T_1, T_2, \ldots, T_t\}\). For an EV, \(EV_k \in \mathbb{K}\), the SoC value on day \(d \in D\) at time \(t \in T\) is denoted by \(R_k(d, t)\). At any given time \(t\),  \(EV_k\) sends a charging request to the CC containing information such as the SoC (\(R_k(d, t)\)) which contains the battery $SoC_k$ and the estimated  $TCC_k$. A $SoC_k$ of $1$ indicates that the battery is fully charged, while a $SoC_k$ of 0 indicates that the battery is completely discharged. The $TCC_k$ refers to the designated period at which the charging request is no longer valid, indicating that the EV must commence charging before the $TCC_k$ expires. Once the CC has received the charging requests from EV ($R_k(d, t)$) within a specific region, it can allow all EVs to charge if the cumulative charging demand does not surpass the maximum capacity for charging energy ($E$). Alternatively, the CC must implement a charging coordination mechanism to choose a subset of EVs for charging within the present time interval while postponing the charging of the remaining EVs to subsequent time intervals. The proposed mechanism establishes a prioritization system for requests, ensuring that the highest-priority requests ${\omega}_k$ are selected for charging while adhering to the constraint of not exceeding the charging energy limit.
\begin{equation} {\omega}_k = \upsilon \times EV_{1}({SoC}_{k}) + (1 - \upsilon  ) EV_{2}({TCC}_{k}), \end{equation} 
where $0\leq\upsilon \leq1$, $EV_1$, and $EV_2$ are functions that ensure the  $SoC_k$ and $TCC_k$ values between 0 and 1. The energy demand of each $EV_k$ ($\theta_k=(1 - SoC_k)B$) is divided by ${\omega}_k$ where $B$ is the overall battery capacity to select the $EVs$ with a higher value of charging request that ensures the maximum charging capacity ($P$) is not surpassed: 
\begin{equation}
\sum \limits _{k \in \partial} \theta _{k} \leq P;~~\text {where}~~\partial \subseteq \mathcal{K}\label{equ:limit}
\end{equation} 
The set $\partial$ represents a subset of $\mathcal{K}$ with a higher priority and adheres to the constraint specified in equation \eqref{equ:limit}. As a result, the CC enables every $EV_k$ to either charge according to its energy requirement $\omega_k$ or postpone the request for a later time slot. 

Due to the absence of sophisticated detection measures, a malicious EV can provide inaccurate SoC information to secure more priority and energy allocation. As a result, EVs that possess malicious intent and rudimentary false reporting capabilities have the potential to engage in power theft, disrupt charging schedules, and undermine the stability of the electrical grid. Thus, the main problem is how to develop and build a robust DL-based IDS at the charging station to prevent any malicious activity, considering the absence of available datasets or incident reports of such an attack as well as the availability of advanced DL architectures that enable the malicious EV to launch successful attacks by adding optimized perturbations using DL models.    \vspace{-5pt}


\section{Proposed Solution}
\label{sec:proposed_solution}
\vspace{-3pt}

This section presents our proposed approach and provides details about the methodology used.
Our hierarchical adversarial DRL approach,  consists of two schemes. The first scheme develops intelligent, covert attacks to falsify SoC data to optimize the charging power amount and prioritize the malicious EV, which enhances the sophistication of the generated dataset. The second scheme is a deep DRL-based IDS for charging stations.
This hierarchical structure has multiple advantages. First, by utilizing the LSTM or Transformers, we can create complex attack strategies. Second, using the sophisticated dataset generated improves the detection model's robustness against various deceptions. The DRL agent is fine-tuned to evade detection, requiring minimal data for attack generation, unlike GANs, which need large datasets. This DRL-based approach allows for autonomous self-learning, bypassing the need for human-driven supervised learning.
\vspace{-5pt}
\subsection{Adversarial DRL Agent Scheme}
\vspace{-3pt}
One fundamental aspect of constructing effective IDSs is the use of robust datasets for training. Our proposed adversarial DRL agent creates more complex and reliable datasets by using LSTM and Transformers as DRL model architectures, rather than just handcrafted or simple DL architectures.
Our approach frames the SoC attack scenario as a Markov Decision Process (MDP), allowing for modeling of attack dynamics within the DRL environment. Key components of this model include \textbf{State ($s_t$):} Defined as the aggregate of incoming charging requests from EVs at time $t$, represented by vector \( \mathbf{q} \) with elements \( {q}_i\) indicating SoC values \( {{SoC}_i} \).
\textbf{Action:} The agent selects a normalized value to distort the SoC of an adversary EV at each $t$, represented by \( \phi_i(t) = S_i(t) + o_t \), where $\mathbf{o_t \in [-1,1] }$. \textbf{Reward function:} Calculated as the cumulative sum of power allocated to the malicious EV, reduced by a penalty for significant perturbations defined as:
\begin{equation}
    \mathbf{w} = \sum_{t=1}^{T} \left( C_i(t) - \nu a_t \right),
    \label {equ:4}
\end{equation}
where $\nu a_t$ is an intrinsic reward to encourage the agent to be stealthy. 
\textbf{ policy:} \( \pi \) comprises a series of actions corresponding to SoC value adjustments. The policy network, based on observed SoC values, formulates a probabilistic policy \( p_{\Theta}(\pi | S) \), utilizing model parameters \( \Theta \).
The DRL agent balances maximizing power gain with maintaining stealth, using a hyperparameter \( \nu \) to modulate the reward function. 
For the models, the LSTM DRL model contains an input layer and three hidden layers of 236 neurons, followed by a ReLU activation function. The output layer, consisting of two neurons, represents a normal distribution's mean \( \mu \) and standard deviation \( \sigma \). The action \( a_t \) is drawn from this distribution. We utilized a BERT (Bidirectional et al. from Transformers) framework configured via BertConfig to process sequential input data in a DRL context. The model employs a Transformers architecture comprising hidden layers and attention heads to capture bidirectional contextual information. The transformer's output is further processed through linear layers, incorporating ReLU activation, culminating in a two-dimensional output space. Additionally, a sampling method is implemented to generate probabilistic outputs, leveraging the normal distribution.
To stabilize the DRL algorithm, we incorporate a baseline term \( b(s) \), calculated as the exponential moving average of the initial loss function \( L(\pi) \) and updated with a decay rate \( \beta \). 

\begin{algorithm}
\caption{Stealthy DRL Agent Training}
\begin{algorithmic}[1]
\State \textbf{Input:} $\rho_e$, $CC$, $N$,$\mathcal{M}$, $\beta$,  $P$.
\State \textbf{Output:} Perturbed Dataset
\State Initialize $Slots = 48$, $EVs = 30$.
\For{j = $0$ \textbf{to} $N$ epochs}
    \State $Set$ $w$ = $0$
    \State $S_j \rightarrow \{M\}$
    \For{$t \rightarrow 1$ \textbf{to} $T$}
        \State $r \sim \Lambda(\beta)$
        \State $Soc_n(t) \sim U[0, 1]$
        \State $ \phi_j \rightarrow Soc_n(t)(t) \cup Soc_j(t)$
        \State $(\mu, \sigma) \rightarrow ( $p$_{\Theta}(\phi_j(t))$
        \State $o_t \sim N(\mu, \sigma)$
        \State $\psi(t) \rightarrow S_b(t) \cup (S_j(t) + o_t)$
        \State $\theta(T)  \rightarrow (1 - \psi(t))B$
        \State $k(t) \rightarrow CC(\theta(t))$
        \State $w \rightarrow  w + C_j(t) - \nu a_t$
    \EndFor
    \State $\Theta \rightarrow \Theta - \iota \nabla L(\Theta|S)$
\EndFor
\end{algorithmic}

\end{algorithm}
Algorithm 1 shows an overview of the training process for the Stealthy DRL agent. We assume several EVs send charging requests to CC each day (Episode). To simulate the real-life scenario, we use a Poisson distribution for the charging request sent to the CC $r$ from the training dataset M, using the arriving rate $\beta$ and $\iota$ as the learning rate, and then $r$ benign SoC values are randomly selected from a uniform distribution. The malicious request of the adversary EV is denoted as $SoC_j$  while $SoC_n$ defines the SoC values of the non-mediocre EVs.

\begin{algorithm}
\caption{DRL-IDS algorthim}
\label{algo:IDS}
\begin{algorithmic}[1]
\State {\textbf{Input:}  $E ,\gamma,\lambda,\epsilon,\alpha ,\theta,\phi$,$B_{\textrm{size}},M$.}

\State \textbf{Output:}  DRL-BASED IDS

\State Read the dataset generated in Algorithm 1
\State Initialize policy network \( \pi_{\theta}(a|s) \) with parameters \( \theta \)
\State Initialize value function network \( V_{\phi}(s) \) with parameters \( \phi \)
\State Initialize replay buffer \( B \)
\For{\( e = 1 \) to \( M \)}
    \State \( D_{\pi_{\theta}}^e \leftarrow \) collect trajectories from \( E \)
    \State \( A_{\theta}^e \leftarrow \) compute advantages using GAE
    \For{each mini-batch in \( D_{\pi_{\theta}}^e \)}
        \State \( (s, a, r, s', y) \leftarrow \) extract mini-batch
        \State \( P(a|s;\theta) \leftarrow \) predict action probability
        \State \( r \leftarrow \) assign reward based on \( P \) and \( y \)
        \State store \( (s, a, r, s') \) in \( B \)
        \State \( \theta \leftarrow \) optimize policy parameters
        \State \( \phi \leftarrow \) optimize value function parameters
    \EndFor
\EndFor
\State \Return \( \theta^* \)
\end{algorithmic}
\end{algorithm}
\subsection{IDS Architecture}
\vspace{-5pt}
This section introduces our proposed methodology for developing a DRL-based IDS. 
As stated before, we develop this IDS based on the datasets generated using our DRL agent. 
A specific encoding approach is utilized to adapt this dataset to a DRL framework, leveraging Mini-Batch SGD. This approach encompasses the treatment of all features, except the `Label, as the current state $S_t$, the `Label` feature itself as the current action $A_t$, and all features in subsequent dataset entries, excluding `Label, as the next state $S_{t+1}$. This method effectively transforms the dataset into a format amenable to DRL, resulting in input data tuples \([S_t, A_t, S_{t+1}]\). 
The Mini-Batch module plays a crucial role, operating as a stochastic subset selector from the original dataset, ensuring non-repetitive sampling throughout each training cycle until the entire dataset is traversed.  
Within this setup, the input to the DRL module consists of sequences of \(n+1\) tuples, each formatted as \([S_t, A_t, S_{t+1}]\), which compose each training batch. The detailed components are described as:\\
\textbf{Environment:} 
The environment for this DRL-based IDS is the one where the dataset generated by stealthy DRL agents is employed.  The features of the generated dataset denote the states of the model. There are 49 features in the generated dataset, and we use the 48 features as states. Feature 49 is the label that will be used for computing the award vectors based on model prediction. In this DRL IDS model, the agent only gets actions to calculate the reward values, and no real action is conducted on the environment.
\textbf{State:}  The states represent the environmental inputs that are received by an agent to perform actions within the framework of DRL. The generated dataset consists of 49 features, from which a specific minibatch is chosen for training purposes by utilizing DRL techniques. \textbf{Rewards:} The reward function, denoted as $R(s_t,a_t,s_{t+1},y),$ reflects the degree of alignment between the current state and the true class label. 
Formally, the rewards for this model are determined by:
\begin{itemize}
    \item +1 if the agent accurately identifies the attack.
\item 0 normal
\item -1 if the agent fails to generate a warning in the event of an attack or if the agent generates an alert for benign.
\end{itemize}
\textbf{Policy} The process of associating an agent's current state with appropriate actions.\\
\textbf{Value} The transition probability, represented as \( P_a(s, s') \), quantifies the likelihood that action \( a \) taken in state \( s \) at time \( t \) will result in state \( s' \) at time \( t+1 \). This can be mathematically expressed as:
\[ P_a(s, s') = \Pr(s' \mid s, a) \]
We utilized the  Proximal Policy Optimization (PPO2), which is based on the actor-critic (AC) algorithm. It utilizes a neural network policy to predict actions based on network states, which are processed to classify fed dataset data as either normal charging requests or potentially intrusive charging requests. The training algorithm procedure is illustrated in Algorithm \ref{algo:IDS}. 
The policy network is represented as $\pi_{\theta'}(a_t|s_t)$, with $\theta'$ denoting the updated policy parameters. The objective function for the policy network is given by:\begin{equation}
J(\theta) = \mathbb{E}_t\left[\min\left(\rho_t(\theta) \tilde{A}_t, \text{clip}\left(\rho_t(\theta), 1 - \epsilon, 1 + \epsilon\right) \tilde{A}_t\right)\right],
\end{equation}
where $\rho_t(\theta) = \frac{\pi_{\theta}(a_t|s_t)}{\pi_{\theta'}(a_t|s_t)}$ represents the ratio of probabilities from the trained policy to the sampling policy. The clipping parameter is represented by $\epsilon$, and $\tilde{A}_t$ is the advantage estimator. The advantage estimator $\tilde{A}_t$ is calculated as:
\begin{equation}
\tilde{A}_t = \delta_t + \beta \delta_{t+1} + \cdots + \beta^{T-t} \delta_T,
\end{equation}
where $\beta$ is the discount factor, $\delta_t = r_t + \beta V_{\phi}(s_{t+1}) - V_{\phi}(s_t)$, with $r_t$ denoting the reward at time step $t$.
Additionally, the loss function for the critic network, which updates the value function parameters $\phi$, is given by:
\begin{equation}
J(\phi) = \mathbb{E}_t[\delta_t^2],
\end{equation}
where $\delta_t^2$ is the squared temporal difference error. This function aims to improve the accuracy of the value function's predictions, enhancing the IDS agent's ability to classify network behaviors effectively.

\begin{figure*}[t]
  \centering
  \begin{subfigure}[b]{0.3\linewidth}
    \includegraphics[width=\linewidth]{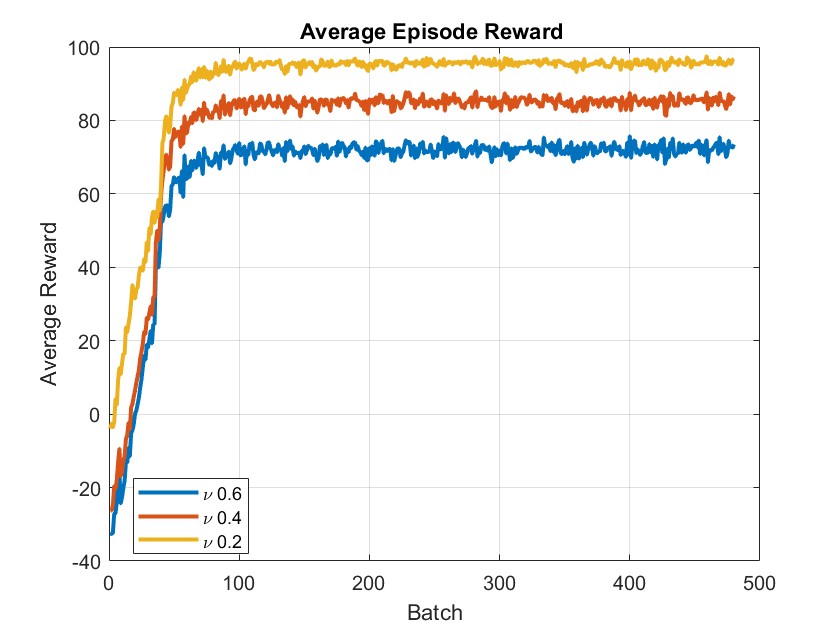}
    \caption{}
    \label{fig:lstm_con}
  \end{subfigure}
  \hspace{0.4cm}
  \begin{subfigure}[b]{0.3\linewidth}
    \includegraphics[width=\linewidth]{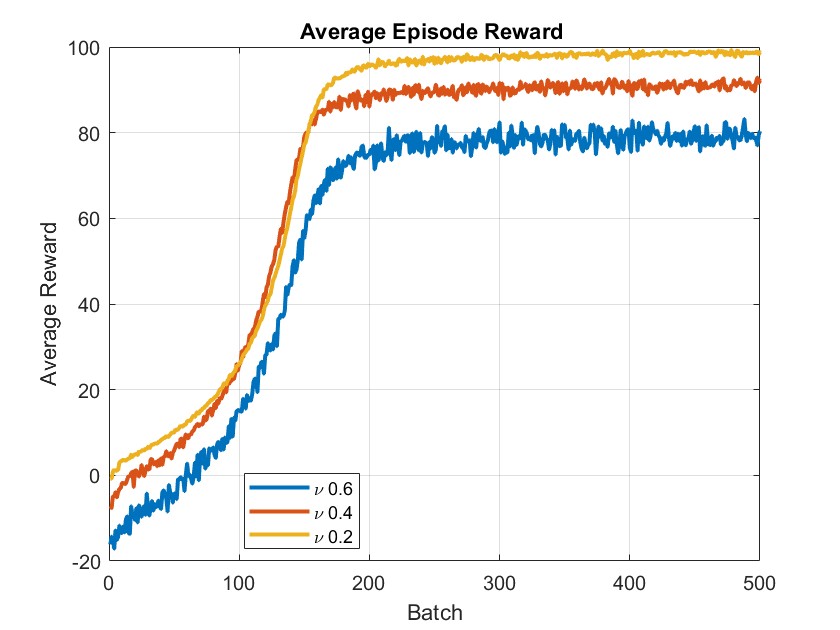}
    \caption{}
    \label{fig:Trans_conv}
  \end{subfigure}
\vspace{-5pt}
  \caption{
Convergence of the stealthy DRL agent with various random seeds.(a) LSTM-based DRL Model (b) Transformers-based DRL Model.}
  \label{fig:mal_conve}
   
\end{figure*}
\vspace{-10pt}

\begin{figure*}[t]
  \centering

  \begin{subfigure}[b]{0.3\linewidth}
    \includegraphics[width=\linewidth]{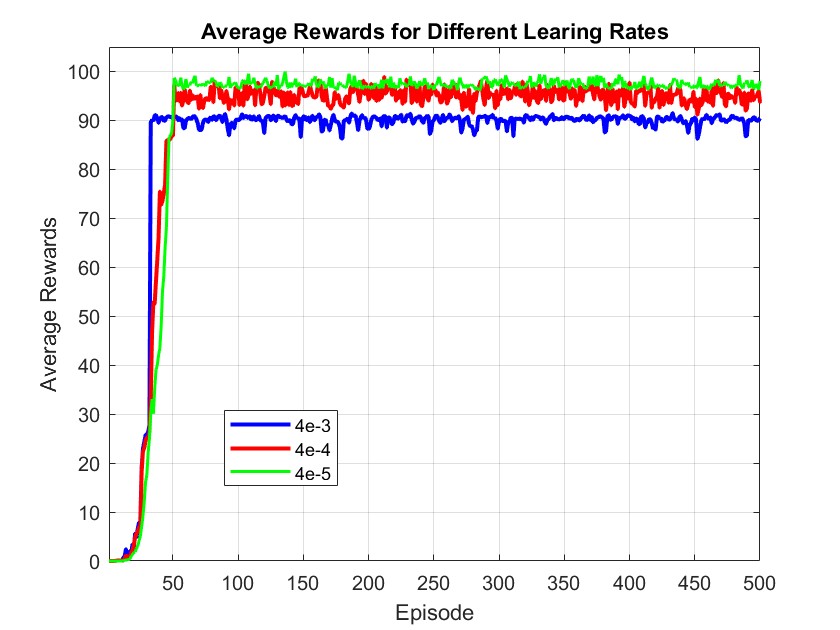}
    \caption{ }
    \label{fig:lstm_con}
  \end{subfigure}
  \hspace{0.4cm}
  \begin{subfigure}[b]{0.3\linewidth}
    \includegraphics[width=\linewidth]{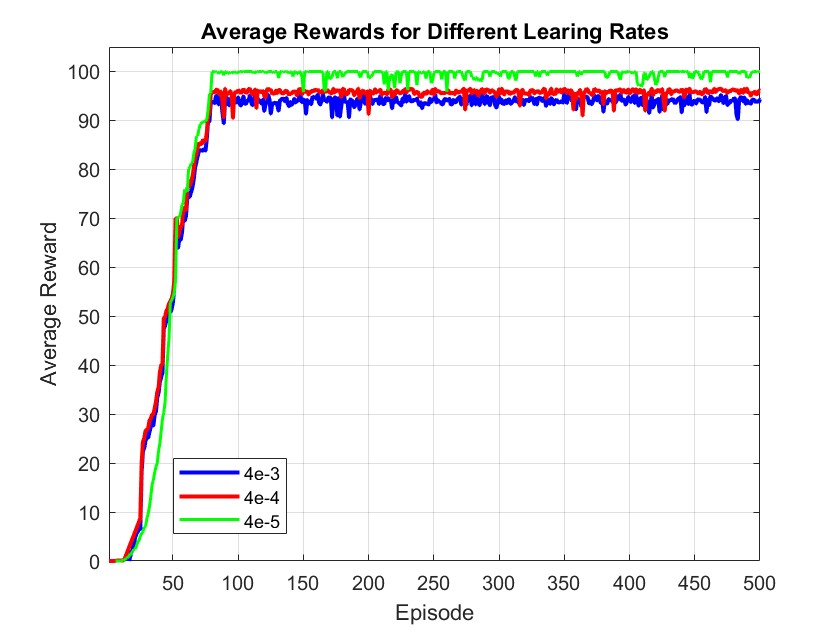}
    \caption{}
    \label{fig:Trans_conv}
  \end{subfigure}
 \vspace{-5pt}
  \caption{
DRL-IDS convergence ((a) LSTM-Based model, (b) Transformers-based model) for different Learning rate values. }
  \label{fig:mal_conve}
\end{figure*}


\section{Performance Evaluation}
\label{sec:results}
In this section, we aim to show the effectiveness of our proposed approach in detecting denial-of-charge attacks against EVCSs. 
\vspace{-3pt}
\subsection{Experimental Setup}
\vspace{-3pt}
We conduct all experiments on a workstation with an Intel I7-10750CPU @ 2.6 GHz processor and 32 GB of memory. The models are trained and tested on a single NVIDIA GeForce GTX 1660 and implemented using PyTorch 1.13.1. These models undergo training utilizing the Adam optimization algorithm. Hyperparameters, including the learning rate ($\alpha$) and the discount factor, are meticulously calibrated on the validation dataset using a grid search strategy.

We employ a dataset comprising 536 Plugin Hybrid Electric Vehicles (taxis) \cite{akhavan2014developing}. These vehicles consistently provided information on their geographic coordinates (latitude and longitude) every minute, as well as their charging duration over 24 days. Additionally, it is assumed that the dataset pertains to the Kia Soul EV \cite{EV}. To estimate the SoC values on a minute-by-minute basis from the driving traces, the charging rates and battery capacity provided by Kia are utilized. The SoC value is updated during the process of charging or driving through the utilization of the following equations
$\text{SoC} =SoC + \frac{\text{Charging level} \times \text{Time}}{\text{Battery size},}$ and
$\text{SoC} =SoC - \frac{\text{Expenditure rate} \times \text{Time}}{\text{Battery size}}.$

To generate a data tuple, the SoC value is sampled at regular intervals of 30 minutes, resulting in a sequence of 48 SoC values over a single day. The total number of data samples obtained is 12,864, which is computed by multiplying the number of taxis (536) by the duration in days (24). The distribution of SoC for two taxis was observed over 23 days.  To obtain a novel attack dataset, we iteratively employ the trained DRL model to create advanced and intelligent stealthy attacks based on each data tuple. For each data pair in the benign dataset, the DRL model is used in the charging simulation to change the actual state of the SoC values of the malicious EVs. The flustered sequence that emerges is designated as a malicious data tuple.  The ADASYN method \cite{he2008adasyn} is subsequently employed as a data augmentation method to achieve a balanced ratio between benign and maliciously generated data. 
\vspace{-0.65em}
\subsection{Results and Discussion}
\vspace{-3pt}
\cref{fig:mal_conve} illustrates the average reward achieved per episode during the process of $500$ epochs of training. The models exhibit high resilience across many random iterations and achieve rapid convergence. We train two agents with $\nu$ values ranging from $0.2, 0.4$, and  $0.6$. It is evident that as the value of $\nu$ increases, the agents' convergence to a lower reward. This arises from the compromise between preserving stealth and gaining more charging power and precedence in the charging schedules. Increasing the value of $\nu$ compels the agent to cause fewer disturbances while acquiring greater control than benign EVs. We notice the LSTM architecture shows fast convergence, while the Transformers architecture shows better performance.

To assess the efficacy of the intelligent and stealthy attacks (i.e., the second scheme), we train the DRL-based IDS model using the generated datasets containing varying combinations of attacks and benign data. Additionally, to study the robustness of the DRL IDS, we trained it with different learning rates ($\alpha = $ $4e-3$, $4e-4$, $4e-5$).
Initially, we trained the model with a learning rate of \(\alpha = 4e-3\), as shown in \cref{fig:mal_conve} by the blue line. This scenario proved the least effective for both architectures (a and b), as the model converged to the lowest reward value, indicating the lowest accuracy. The figure also illustrates the training process at a learning rate of \(4e-4\). It is evident from the figure that the model becomes unstable at this particular learning rate, with the reward reaching a convergence point of \(0.90\) for the LSTM and \(0.94\) for the Transformers and then experiencing expected fluctuations due to the learning process. Finally, as depicted in the accompanying figure, we trained the model using a learning rate of \(4e-5\), resulting in the highest accuracy for both models. The remaining simulations approached the maximum.
\begin{table}[]
\centering
\caption{Comparison of performance of DRL-Based IDS using different datasets.}
\label{tab:IDS_results}
\resizebox{\columnwidth}{!}{%
\begin{tabular}{ccccc}
\hline
Dataset                 & Accuracy & Recall & Precision & F1-Score \\ \hline
Proposed LSTM-based     & 0.994    & 0.996  & 0.994     & 0.996    \\
Proposed   Transformers & 0.999    & 1.0    & 0.999     & 0.999    \\
DRL                  & 0.990    & 0.992  & 0.990     & 0.993    \\ \hline
\end{tabular}%
}

\end{table}
Table \ref{tab:IDS_results} reveals further performance differences among the three IDS models. The LSTM-based DRL model showed high effectiveness with an accuracy of 0.994 and an F1-score of 0.996. However, the Transformer-based DRL model excelled further, achieving near-perfect metrics, notably a 1.0 recall, 0.999 accuracy, and F1-score. This superiority is due to the Transformer's advanced pattern recognition capabilities. The more sophisticated LSTM and Transformer models slightly outperformed the DRL model, which was effective with an accuracy of 0.990 and an F1-score of 0.993, with the latter proving to be the most effective for IDS dataset generation.

\section{Conclusion}
\label{sec:concl}
This paper addressed the critical challenge of securing EVCS against sophisticated cyber threats. We developed a framework comprising two key components: the generation of complex synthetic attacks using DRL, and the development of a robust DRL-based IDS. Our approach significantly advances existing security measures by simulating more realistic and challenging cyberattack scenarios, thereby enhancing the resilience of IDS against advanced threats.  We carried out extensive simulations, and our results demonstrated the efficacy of advanced DRL techniques in both crafting intricate cyberattacks and strengthening detection systems. We also tested the detector's ability to detect new attacks not in the training dataset. Our evaluation showed that the detector detected these new attacks well.
\vspace{-3pt}
\section*{Acknowledgement}
\vspace{-3pt}
{
This publication was made possible by TUBITAK-QNRF joint Funding Program (Tubitak-QNRF 4th Cycle grant \# AICC04-0812-210017) from the Qatar National Research Fund (a member of Qatar Foundation). The findings herein reflect the work, and are solely the responsibility of the authors.}
\vspace{-3pt}
\bibliographystyle{IEEEtran}
\bibliography{ref}

\end{document}